\documentclass{svmult}

\usepackage{amsmath}
\usepackage{amssymb}
\usepackage{longtable}
\usepackage{dcolumn}
\usepackage[square,comma,numbers,sort&compress]{natbib}

\newcolumntype{b}{D{(}{\ (}{-1}}
\newcolumntype{d}{D{.}{.}{-1}}
\renewcommand{\textsc}[1]{$\,${\footnotesize #1}}

\def\etal{\emph{et~al.}}

\newcommand{\tref}[1]{Table~\ref{#1}}

\newcommand{\cm}{cm$^{-1}$}

\begin{document}

\title*{Atomic transition frequencies, isotope shifts, and sensitivity to variation of the fine structure constant for studies of quasar absorption spectra.}

\titlerunning{Atomic data for $\alpha$-variation studies of QSO spectra}
\authorrunning{J. C. Berengut \etal}

\author{J. C. Berengut \and V. A. Dzuba \and V. V. Flambaum \and J. A. King
\and M. G. Kozlov \and M. T. Murphy \and J. K. Webb}

\institute{J. C. Berengut \and V. A. Dzuba \and V. V. Flambaum \and J. A. King \and J. K. Webb \at School of Physics, University of New South Wales, Sydney, NSW 2052, Australia
\and
M. G. Kozlov \at Petersburg Nuclear Physics Institute, Gatchina, 188300, Russia
\and
M. T. Murphy \at Centre for Astrophysics and Supercomputing, Swinburne University of Technology, Victoria 3122, Australia
}

\maketitle

\abstract{
Theories unifying gravity with other interactions suggest spatial and temporal variation of fundamental ``constants'' in the Universe. A change in the fine structure constant, $\alpha = e^2/\hbar c$, could be detected via shifts in the frequencies of atomic transitions in quasar absorption systems. Recent studies using 140 absorption systems from the Keck telescope and 153 from the Very Large Telescope, suggest that $\alpha$ varies spatially~\cite{webb10arxiv}. That is, in one direction on the sky $\alpha$ seems to have been smaller at the time of absorption, while in the opposite direction it seems to have been larger.
\newline\indent
To continue this study we need accurate laboratory measurements of atomic transition frequencies. The aim of this paper is to provide a compilation of transitions of importance to the search for $\alpha$ variation. They are E1 transitions to the ground state in several different atoms and ions, with wavelengths ranging from around 900 -- 6000~\AA, and require an accuracy of better than 10$^{-4}$~\AA. We discuss isotope shift measurements that are needed in order to resolve systematic effects in the study. The coefficients of sensitivity to $\alpha$-variation ($q$) are also presented.
}


\section{Introduction}

Current theories that seek to unify gravity with the other fundamental interactions suggest that spatial and temporal variation of fundamental constants is a possibility, or even a necessity, in an expanding Universe (see, for example the review of~\citet{uzan03rmp}). Several studies have tried to probe the values of constants at earlier stages in the evolution of the Universe, using tools such as big-bang nucleosynthesis, the Oklo natural nuclear reactor, quasar absorption spectra, and atomic clocks (see, e.g.~\citet{flambaum09ijmpa}).

Comparison of atomic transition frequencies on Earth and in quasar (QSO) absorption spectra can be used to measure variation of the fine-structure constant $\alpha=e^2/\hbar c$ over the last 10~billion years or so. Early studies used the ``alkali-doublet'' method \citep{savedoff56nat}, taking advantage of the simple $\alpha$-dependence of the separation of a fine-structure multiplet.

More recently we developed the ``many-multiplet'' method \citep{dzuba99prl,dzuba99pra} which improves sensitivity to variation in $\alpha$ by more than an order of magnitude compared to the alkali-doublet method. Enhancement comes from the use of transitions which are more sensitive to $\alpha$ than the fine-structure splitting is, for example the $s$-wave orbital has maximum relativistic corrections to energy but no spin-orbit splitting. In addition the $\alpha$-dependence varies strongly between different atoms and transitions (for example $s$--$p$ and $s$--$d$ transitions can have different signs) and this helps to control instrumental and astrophysical systematics. The number of spectral lines available for study is quite large; this gives a statistical advantage.

The first analyses using the many-multiplet method and quasar absorption spectra obtained at the Keck telescope revealed hints that the fine structure constant was smaller in the early universe \citep{webb99prl,webb01prl,murphy01mnrasA,murphy01mnrasC,webb03ass,murphy03mnras,murphy04lnp}. A very extensive search for possible systematic errors has shown that known systematic effects cannot explain the result~\citep{murphy01mnrasB}.

Our method and calculations have been used by other groups to analyse a different data set from the Very Large Telescope (VLT) in Chile~(\citet{srianand04prl}), and their results indicate no variation of $\alpha$ (see also~\cite{chand04aap}). It was noted later that there were sharp fluctuations in chi-squared vs. $\Delta\alpha/\alpha$ graphs of~\cite{srianand04prl,chand04aap} that indicate failings in the chi-squared minimisation routine~\cite{murphy07prl}, and it was shown that the errors were underestimated by a large factor~\cite{murphy07prl,murphy08mnras,srianand07prl}.

A large scale analysis, combining the Keck data with a new sample of 153 measurements from the VLT, indicates a spatial variation in $\alpha$ at the 4.1$\sigma$ level~\cite{webb10arxiv}. This gradient has a declination of around $-60^\circ$, which explains why the Keck data, restricted mainly to the northern sky since the telescope is in Hawaii at a latitude of 20$^\circ$~N, originally suggested a time-varying $\alpha$ that was smaller in the past. The VLT is in Chile, at latitude $25^\circ$~S, giving the new combined study much more complete sky coverage. The new results are entirely consistent with previous ones. Other results from other groups using single ions in single absorption systems~\cite{quast04aap,levshakov05aap,levshakov06aap,levshakov07aap} are also consistent with the dipole result~\cite{berengut10arxiv2}. We note that individual sight-lines are inherently less useful than large samples, no matter what the signal-to-noise ratio of the single sight-line spectra, because some systematic errors that are present for single sight-lines often randomize over a large sample~\cite{murphy01mnrasB}.

\section{Laboratory spectroscopy}

To continue this work, several new transitions are being considered. In \tref{tab:wavelengths} we present a list of lines commonly observed in high-resolution QSO spectra. All of the lines marked `A' (extremely important), `B' (very important), or `C' (less important) lack the high-accuracy laboratory measurements necessary for studies of $\alpha$ variation. All transitions are from the ground state of the ion, with the exception of the C\textsc{II} lines marked with an asterisk which are transitions from the metastable $2s^22p\ ^2P_{3/2}^o$ level. Predominantly the wavelengths and oscillator strengths are taken from the compilations of \citet{morton91apjss,morton03apjss}. The wavelengths have errors of about 0.005~\AA, although it is possible that some errors are closer to 0.05~\AA. Note that the oscillator strengths presented are not as accurate as the wavelengths: these measurements are much more difficult. As a general rule, the lines are more important for $\alpha$ variation if they lie above 1215.67~\AA\ (the Lyman-$\alpha$ line of hydrogen) due to the ``Lyman-$\alpha$ forest'' seen in QSO spectra.

Isotope shift measurements for these transitions are also needed in order to resolve a possible source of systematic error in the variation of $\alpha$ studies: the isotope abundance ratios in the gas clouds sampled in the quasar absorption spectra may not match those on Earth~\citep{murphy01mnrasB,murphy03ass}. Spurious observation of $\alpha$-variation due to differences in isotope abundance of any one element has been ruled out~(see, e.g.~\cite{murphy03mnras}), however an improbable ``conspiracy'' of changes in several elements could mimic an effect. On the other hand, it is difficult to see how such changes could lead to spurious observation of a spatial variation since the underlying mechanisms of chemical evolution would have to vary spatially. Nevertheless, the many-multiplet method uses different transitions of different atoms at different redshifts, so ignoring the isotopic structure of transitions may destroy the consistency between sub-samples occupying different redshift ranges.

Accurate measurements of the isotope shift are required to quantify these systematic effects. Additionally, if the isotope shifts are known then it is possible to simultaneously determine both any possible $\alpha$-variation and the isotope abundances in the early universe directly \citep{kozlov04pra}. This can be used to constrain models of chemical evolution of the Universe and test models of nuclear processes in stars \citep{ashenfelter04prl,fenner05mnras}. We have performed very complicated calculations of some isotope shifts~\citep{berengut03pra,berengut05pra,berengut06pra,berengut08jpb}, however calculations in group $3d$ atoms and ions are difficult, and our accuracy may be low. Therefore measurements for at least some lines are needed to benchmark calculations in this regime. In \tref{tab:wavelengths} we indicate lines for which isotope shifts are known by a \checkmark\ in the `I.S.' column. Lines that were used in previous studies (and hence have precise wavelength measurements), but for which the isotopic structure has not been measured are marked with an `A' in this column.

A similar systematic effect to that caused by isotope abundances can occur due to differential saturation of the hyperfine components. This can occur because of deviations from local thermal equilibrium (see, e.g.~\cite{murphy03mnras}). While we do not discuss this issue in this paper, we note that in some cases, such as Al\,III and Mn\,II, hyperfine structure can be as important as isotopic structure.

\section{Sensitivity coefficients}

We previously calculated the relativistic energy shifts, or $q$-values, for many of the lines seen in quasar spectra \citep{dzuba02praA,berengut04praB,berengut05pra,dzuba05pra,berengut06pra,dzuba07pra,savukov08pra,porsev07pra}. The difference between the transition frequencies in QSO spectra ($\omega$) and in the laboratory ($\omega_0$) depends on the relative values of $\alpha$. The dependence of the frequencies on small changes in $\alpha$ is given by the formula 
\begin{gather}
\omega=\omega_0+qx \,,\\
\nonumber
x=(\alpha/\alpha_0)^2 -1 \approx 2\frac{\alpha - \alpha_0}{\alpha_0} \,.
\end{gather}
The $q$ values are calculated using atomic physics codes. The atomic energy levels are calculated to a first approximation using relativistic Hartree-Fock (Dirac-Hartree-Fock). Higher order effects are taken into account using a combination of configuration interaction (for many-valence-electron systems) and many-body perturbation theory; this is known as the ``CI+MBPT'' method~\citep{dzuba96pra}. The value of $\alpha$ is varied in the computer codes and the energy levels are recalculated, and hence the transition frequencies. The $q$ values are extracted as
\[
q = \frac{d \omega }{d x} \bigg|_{x=0}
\]
We also account for complications due to level pseudo-crossing as described by~\citet{dzuba02praA}. In \tref{tab:wavelengths} we present our current best $q$-values for easy reference. Note that for Fe\,II lines, we present the arithmetic average of the independent calculations~\cite{dzuba02praA} and~\cite{porsev07pra}. Uncertainties here are representative rather than statistical.

\subsubsection*{Acknowledgements}

The authors would like to thank D.~Morton and W.~Ubachs for useful comments and for pointing out some errors.

\newpage

\begin{longtable}{lcccbccc}
\caption{\label{tab:wavelengths} High-priority lines observed in QSO spectra. The need for precise wavelength measurement (or re-measurement) is indicated by an A (extremely important), B (very important) or C (less important) in the `Status' column; existing precise measurements are referenced in the last column. In the `I.S.' column the status of isotope structure measurement is indicated: a \checkmark\ means it has been measured, and `A' means that it is unknown and urgently needed. All transitions are from the ground state of the ion, with the exception of the C\textsc{II} lines marked with an asterisk which are transitions from the metastable $2s^22p\ ^2P_{3/2}^o$ level.
}\\
\hline \hline
Atom/ & Wavelength     &  Frequency      & Oscillator & \multicolumn{1}{c}{$q$ value} & \multicolumn{2}{c}{Status} & Refs. \\
Ion   & $\lambda$ (\AA)& $\omega_0$ (\cm)& Strength   & \multicolumn{1}{c}{(\cm)}     & $\omega_0$ & I.S. & \\
\hline \hline \endfirsthead
\caption{(continued)} \\
\hline \hline
Atom/ & Wavelength     &  Frequency      & Oscillator & \multicolumn{1}{c}{$q$ value} & \multicolumn{2}{c}{Status} & Refs. \\
Ion   & $\lambda$ (\AA)& $\omega_0$ (\cm)& Strength   & \multicolumn{1}{c}{(\cm)}     & $\omega_0$ & I.S. & \\
\hline \hline \endhead
\hline \endfoot
\hline \hline \endlastfoot
C\,I
    &  945.188 & 105799.1 & 0.272600 & 130(60) & C & \checkmark & \cite{labazan05pra} \\*
    & 1139.793 & 87735.30 & 0.013960 & 0(100) & C & \\*
    & 1155.809 & 86519.47 & 0.017250 & 0(100) & C & \\*
    & 1157.186 & 86416.55 & 0.549500 & 0(100) & C & \\*
    & 1157.910 & 86362.52 & 0.021780 & 0(100) & C & \\*
    & 1188.833 & 84116.09 & 0.016760 & 0(100) & C & \\*
    & 1193.031 & 83820.13 & 0.044470 & 0(100) & C & \\*
    & 1193.996 & 83752.41 & 0.009407 & 0(100) & C & \\*
    & 1260.736 & 79318.78 & 0.039370 & 30(10) & B & \\*
    & 1276.483 & 78340.28 & 0.004502 & 17(10) & B & \\*
    & 1277.245 & 78293.49 & 0.096650 & -13(10)& B & \\*
    & 1280.135 & 78116.74 & 0.024320 & -21(10)& B & \\*
    & 1328.833 & 75253.97 & 0.058040 & 117(10)& B & \\*
    & 1560.309 & 64089.85 & 0.080410 & 137(10)& B & \\*
    & 1656.928 & 60352.63 & 0.140500 & -24(10)& B & \\
C\,II
    & 1036.337 & 96493.74 & 0.123000 & 168(10) & B & \\*
    & 1037.018$^*$ & 96430.32 & 0.123000 & 105(10) & B & \\*
    & 1334.532 & 74932.62 & 0.127800 & 178(10) & B & \\*
    & 1335.662$^*$ & 74869.20 & 0.012770 & 115(10) & B & \\*
    & 1335.707$^*$ & 74866.68 & 0.114900 & 118(10) & B & \\
C\,III
    &  977.020 & 102352.0 & 0.762000 & 165(10) & C & \\
C\,IV
    & 1548.204 & 64590.99 & 0.190800 & 222(2) & B & & \cite{griesmann00apj} \\*
    & 1550.781 & 64483.65 & 0.095220 & 115(2) & B & & \cite{griesmann00apj} \\
O\,I
    & 1025.762 & 97488.53 & 0.020300 & 0(20) & C & & \cite{ivanov08mnras} \\*
    & 1026.476 & 97420.72 & 0.002460 & 0(20) & C & \\*
    & 1039.230 & 96225.05 & 0.009197 & 0(20) & C & \\*
    & 1302.168 & 76794.98 & 0.048870 & 0(20) & B & \\
Na\,I 
    & 3303.320 & 30272.58 & 0.013400 & 57(2) & C & \\*
    & 3303.930 & 30266.99 & 0.006700 & 51(2) & C & \\*
    & 5891.583 & 16973.37 & 0.655000 & 62(2) & C & \checkmark & \cite{juncar81met,pescht77zpa,huber78prc} \\*
    & 5897.558 & 16956.17 & 0.327000 & 45(2) & C & \checkmark & \cite{juncar81met,gangrsky98epja} \\
Mg\,I
    & 2026.477 & 49346.73 & 0.112000 & 87(7) & B & \checkmark & \cite{aldenius06mnras,hannemann06pra} \\*
    & 2852.963 & 35051.27 & 0.181000 & 90(10) & B & \checkmark & \cite{aldenius06mnras,pickering98mnras,hallstadius79zpa,boiteux88jpf,salumbides06mnras} \\
Mg\,II
    & 1239.925 & 80650.04 & 0.000267 & 192(2) & C & \\*
    & 2796.354 & 35760.85 & 0.612300 & 212(2) & B & \checkmark & \cite{aldenius06mnras,pickering98mnras,drullinger80ap,batteiger09pra} \\*
    & 2803.532 & 35669.30 & 0.305400 & 121(2) & B & \checkmark & \cite{aldenius06mnras,pickering98mnras,batteiger09pra} \\
Al\,II
    & 1670.789 & 59851.97 & 1.880000 & 270(30) & B & \checkmark & \cite{griesmann00apj} \\
Al\,III
    & 1854.718 & 53916.54 & 0.539000 & 458(6) & B & \checkmark & \cite{griesmann00apj} \\*
    & 1862.791 & 53682.88 & 0.268000 & 224(8) & B & \checkmark & \cite{griesmann00apj} \\
Si\,II
    & 1190.416 & 84004.26 & 0.250200 & & C & \\*
    & 1193.290 & 83801.95 & 0.499100 & & C & \\*
    & 1260.422 & 79338.50 & 1.007000 & & B & \\*
    & 1304.370 & 76665.35 & 0.094000 & & B & \\*
    & 1526.707 & 65500.45 & 0.117094 &  50 (30) & B & A & \cite{griesmann00apj} \\*
    & 1808.013 & 55309.34 & 0.002010 & 520 (30) & B & A & \cite{griesmann00apj} \\
Si\,IV
    & 1393.760 & 71748.64 & 0.528000 & 823(40) & B & A & \cite{griesmann00apj} \\*
    & 1402.773 & 71287.54 & 0.262000 & 361(15) & B & A & \cite{griesmann00apj} \\
S\,II
    & 1250.583 & 79962.61 & 0.005350 & & B & \\*
    & 1253.808 & 79756.83 & 0.010700 & & B & \\*
    & 1259.518 & 79395.39 & 0.015900 & & B & \\
Ca\,II
    & 3934.777 & 25414.40 & 0.688000 & 446(6) & B & \checkmark & \cite{wolf08pra} \\*
    & 3969.591 & 25191.51 & 0.341000 & 222(2) & B & \checkmark & \cite{wolf08pra} \\
Ti\,II
    & 1910.600 & 52339.58 & 0.202000 & -1564(150) & A & A \\*
    & 1910.938 & 52330.32 & 0.098000 & -1783(300) & A & A \\*
    & 3067.245 & 32602.55 & 0.041500 & 791(50) & B & A & \cite{aldenius06mnras} \\*
    & 3073.877 & 32532.21 & 0.104000 & 677(50) & B & A & \cite{aldenius06mnras} \\*
    & 3230.131 & 30958.50 & 0.057300 & 673(50) & B & A & \cite{aldenius06mnras} \\*
    & 3242.929 & 30836.32 & 0.183000 & 541(50) & B & A & \cite{aldenius06mnras} \\*
    & 3384.740 & 29544.37 & 0.282000 & 396(50) & B & A & \cite{aldenius06mnras} \\
Cr\,II
    & 2056.256 & 48632.06 & 0.105000 & -1110 (150) & B & A & \cite{aldenius06mnras,pickering00mnras} \\*
    & 2062.236 & 48491.05 & 0.078000 & -1280 (150) & B & A & \cite{aldenius06mnras,pickering00mnras} \\*
    & 2066.164 & 48398.87 & 0.051500 & -1360 (150) & B & A & \cite{aldenius06mnras,pickering00mnras} \\
Mn\,II
    & 1197.184 & 83529.35 & 0.156600 & -2556(450) & C & \\*
    & 1199.391 & 83375.65 & 0.105900 & -2825(450) & C & \\*
    & 1201.118 & 83255.77 & 0.088090 & -3033(450) & C & \\*
    & 2576.877 & 38806.66 & 0.288000 & 1276(150) & B & A & \cite{aldenius06mnras,blackwell-whitehead05mnras} \\*
    & 2594.499 & 38543.08 & 0.223000 & 1030(150) & B & A & \cite{aldenius06mnras,blackwell-whitehead05mnras} \\*
    & 2606.462 & 38366.18 & 0.158000 &  869(150) & B & A & \cite{aldenius06mnras,blackwell-whitehead05mnras} \\
Fe\,II
    & 1063.176 & 94057.80 & 0.060000 & & C & \\*
    & 1063.971 & 93987.52 & 0.003718 & & C & \\*
    & 1096.877 & 91167.92 & 0.032400 & & C & \\*
    & 1121.975 & 89128.55 & 0.020200 & & C & \\*
    & 1125.448 & 88853.51 & 0.016000 & & C & \\*
    & 1143.226 & 87471.77 & 0.017700 & & C & \\*
    & 1144.939 & 87340.98 & 0.106000 & & C & \\*
    & 1260.533 & 79331.52 & 0.025000 & & B & \\*
    & 1608.450 & 62171.63 & 0.058000 & -1165 (300) & A & A & \cite{pickering02aap} \\*
    & 1611.200 & 62065.53 & 0.001360 & 1330 (300) & A & A & \cite{pickering02aap} \\*
    & 2249.877 & 44446.88 & 0.001821 & & A & A \\*
    & 2260.780 & 44232.51 & 0.002440 & & A & A & \cite{aldenius06mnras}\\*
    & 2344.212 & 42658.24 & 0.114000 & 1375 (300) & B & A & \cite{aldenius06mnras,nave91josab} \\*
    & 2367.589 & 42237.06 & 0.000212 & 1904 & B & \\*
    & 2374.460 & 42114.83 & 0.031300 & 1625 (100) & B & A & \cite{aldenius06mnras,nave91josab} \\*
    & 2382.764 & 41968.06 & 0.320000 & 1505 (100) & B & A & \cite{aldenius06mnras,nave91josab} \\*
    & 2586.649 & 38660.05 & 0.069180 & 1515 (100) & B & A & \cite{aldenius06mnras,nave91josab} \\*
    & 2600.172 & 38458.99 & 0.238780 & 1370 (100) & B & A & \cite{aldenius06mnras,nave91josab} \\
Ni\,II
    & 1317.217 & 75917.64 & 0.146000 & & A & \\*
    & 1370.132 & 72985.67 & 0.076900 & & A & \\*
    & 1393.324 & 71770.82 & 0.022220 & & A & \\*
    & 1454.842 & 68735.99 & 0.032300 & & A & \\*
    & 1467.259 & 68154.29 & 0.009900 & & C & \\*
    & 1467.756 & 68131.22 & 0.006300 & & C & \\*
    & 1502.148 & 66571.34 & 0.006000 & & C & \\*
    & 1703.412 & 58705.71 & 0.012240 &             & A & & \cite{pickering00mnras} \\*
    & 1709.604 & 58493.07 & 0.032400 &   -20 (250) & A & A & \cite{pickering00mnras} \\*
    & 1741.553 & 57420.01 & 0.042700 & -1400 (250) & A & A & \cite{pickering00mnras} \\*
    & 1751.915 & 57080.37 & 0.027700 &  -700 (250) & A & A & \cite{pickering00mnras} \\*
Zn\,II
    & 2026.137 & 49355.00 & 0.489000 &  2470 (25) & C & \checkmark & \cite{aldenius06mnras,pickering00mnras,matsubara03apb} \\*
    & 2062.660 & 48481.08 & 0.256000 &  1560 (25) & B & A & \cite{aldenius06mnras,pickering00mnras} \\*
Ge\,II
    & 1237.059 & 80836.880 & 0.870000 & 2236 (70) & A & \\
    & 1602.486 & 62403.028 & 0.130000 & -664 (70) & B & \\
\end{longtable}

\newpage
\footnotesize
\bibliographystyle{apsrev}
\bibliography{references}

\end{document}